\documentclass[aps,prd,reprint,
amsmath,amssymb]{revtex4-2}

\usepackage{bm}
\usepackage{mathtools}
\usepackage{graphicx}
\usepackage[usenames,dvipsnames]{color}
\usepackage[colorlinks=true,linkcolor=blue]{hyperref}
\usepackage{epsfig}

\newcommand{\tr}{\operatorname{tr}}
\newcommand{\dd}{\mathrm d}
\newcommand{\I}{\mathbb I}
\newcommand{\MG}{\mathcal M_{\rm G}}

\begin{document}

\title{Thermalization-Induced Entropy-Rate Representation of Internal Energy in a Damped Quantum Oscillator}

\author{Hyeong-Chan Kim}
\email{hckim@ut.ac.kr}
\author{Youngone Lee}
\email{samadihi@gmail.com}
\affiliation{School of Liberal Arts and Sciences, Korea National University of Transportation, Chungju 380-702, Korea}

\begin{abstract}
We study a harmonic oscillator weakly coupled to a thermal bath and evolving under the standard quantum-optical master equation.  For zero-mean Gaussian states, entropy alone does not determine the internal energy because the latter also depends on nonequilibrium Gaussian structure.  We show, however, that thermal relaxation yields the exact representation $ E=E(S,\dot S) $ for every mixed Gaussian state evolving under this thermal quantum-optical master equation.  Explicitly,
\[
E(S,\dot S)=
\frac{\omega\nu(S)}{\nu_{\rm th}}
\left[
\nu(S)+
\frac{\dot S}{\gamma S'(\nu(S))}
\right].
\]
Once the thermal GKLS equation is assumed, no further expansion in the damping rate, entropy rate, or distance from equilibrium is made.

The entropy rate is not an alternative instantaneous state coordinate: it depends on the open-system dynamics and therefore carries information about the coupling to the bath.  This is exposed by the isolated limit, where $\dot S=0$ for all Gaussian states and the representation no longer determines the energy.  The result thus identifies an exact thermalization-induced dynamical energetic relation, distinct from rate dependence generated by an external driving protocol.  We also describe an operational test based on time-resolved purity and energy measurements. At fixed frequency the bath only relaxes pre-existing squeezing.  We therefore also consider frequency modulation, which can drive an initially thermal state away from the instantaneous Gibbs family by generating off-Gibbs Gaussian structure.  Within a local Gaussian continuation, the bath then relaxes this driven deformation, providing a dynamical realization of the entropy-rate dependence.
\end{abstract}


\maketitle
\section{Introduction}
\label{sec:intro}

Equilibrium thermodynamics represents the internal energy in terms of
instantaneous state variables, for example \(E=E(S,\lambda)\), with \(S\) the
entropy and \(\lambda\) externally controlled parameters.  Away from
equilibrium, additional state variables are often introduced, while
rate-dependent quantities also arise in finite-time and irreversible
thermodynamics~\cite{JouCasasVazquezLebon2010,MullerRuggeri1998,
JouCasasVazquezLebon1999,Andresen1984,Deffner2010,Salamon1983}.
For open quantum systems, related questions are studied in quantum and
stochastic thermodynamics~\cite{DeffnerCampbell,Kosloff2013,Seifert2012}.
Here we ask a narrower question: can ordinary thermal relaxation itself yield
an exact reduced energetic representation involving the rate of a
thermodynamic variable?

We address this question in the simplest standard setting, a harmonic
oscillator weakly coupled to a thermal bath.  For a zero-mean Gaussian state,
the entropy is determined by the symplectic eigenvalue of the covariance
matrix, whereas the energy also depends on nonequilibrium Gaussian structure.
Entropy alone therefore does not determine the energy away from the Gibbs
family.

For the thermal quantum-optical master equation, the covariance dynamics
provides an additional relation between the energy-carrying second moment and
the entropy rate.  Eliminating that moment gives the exact reduced relation
\begin{equation}
E=E(S,\dot S),
\label{eq:main-schematic}
\end{equation}
valid for every mixed Gaussian state and every nonzero damping rate compatible
with the Born--Markov--secular regime.  No expansion in the damping rate, in
\(\dot S\), or in the distance from equilibrium is required.

The meaning of Eq.~\eqref{eq:main-schematic} is dynamical rather than
kinematic.  The quantity \(\dot S\) is not an additional instantaneous
coordinate of the Gaussian state: its value is fixed only after the
open-system generator has been specified.  Thus the representation
\(E(S,\dot S)\) contains information about thermalization that is absent from
the instantaneous entropy alone.  This distinction becomes sharp in the
isolated limit.  When the damping vanishes, \(\dot S=0\) for every Gaussian
state although equal-entropy states can still have different energies, and
the entropy-rate representation ceases to close.  The derivative dependence
is therefore induced by the bath coupling rather than by the speed of an
externally imposed protocol.

The paper is organized as follows.  Section~\ref{sec:model} specifies the model and assumptions.
We derive this relation directly from the exact second-moment dynamics in Sec.~\ref{sec:moments}, identify its physically accessible domain, and give an operational test based on time-resolved measurements of purity and mean energy, a time-dependent bath temperature in  Sec.~\ref{sec:closure}.  
We then  turn to a driven oscillator, where frequency modulation provides a physical mechanism for generating the off-Gibbs Gaussian structure whose subsequent relaxation is encoded in the entropy rate in Sec.~\ref{sec:driven}.  
The driven construction is formulated as a local Gaussian continuation of the exactly controlled fixed-frequency problem, with its additional assumptions and regime of validity stated explicitly.
Technical details are deferred to the Appendices.

\section{Model, assumptions, and notation}
\label{sec:model}

We set $\hbar=k_B=m=1$.  For an oscillator of fixed frequency $\omega$ we use
the dimensionless quadratures
\begin{equation}
 X=\sqrt{\omega}\,x,
 \qquad
 P=\frac{p}{\sqrt{\omega}},
 \qquad [X,P]=i,
\end{equation}
for which
\begin{equation}
 H=\frac{\omega}{2}(X^2+P^2).
 \label{eq:H-fixed}
\end{equation}
We restrict attention to Gaussian states with vanishing first moments.  Such a
state is specified by the real symmetric covariance matrix
\begin{equation}
 \sigma_{ij}=\frac12\langle\{R_i,R_j\}\rangle,
 \qquad R=(X,P).
\end{equation}
Of its three independent components, two rotational invariants are sufficient
for the energetic and entropic discussion of the isotropic generator used
below~\cite{Weedbrook2012,Williamson1936}:
\begin{equation}
 s=\tr\sigma,
 \qquad
 \nu=\sqrt{\det\sigma}.
 \label{eq:invariants}
\end{equation}
The remaining parameter is the orientation of the covariance ellipse.  It
enters neither the energy nor the entropy and, at constant $\omega$, rotates
independently; Appendix~\ref{app:covariance} gives the component evolution
explicitly.

It is convenient to introduce the anisotropy, or squeezing magnitude,
\begin{equation}
 r^2=\frac14(\sigma_{11}-\sigma_{22})^2+\sigma_{12}^2
     =\frac{s^2}{4}-\nu^2,
 \qquad \nu\geq\frac12.
 \label{eq:nu-s-r}
\end{equation}
The variable $r$ is not independent of $(s,\nu)$.  It is useful because
$r=0$ is equivalent to $\sigma=\nu\I$, namely an unsqueezed Gibbs state of the
fixed oscillator Hamiltonian at some temperature.

The energy and von Neumann entropy are
\begin{align}
 E&=\langle H\rangle=\frac{\omega s}{2},
 \label{eq:E-of-s}\\
 S(\nu)&=\left(\nu+\frac12\right)\ln\left(\nu+\frac12\right)
 -\left(\nu-\frac12\right)\ln\left(\nu-\frac12\right),
 \label{eq:S-of-nu}
\end{align}
with
\begin{equation}
 S'(\nu)=\ln\frac{\nu+\tfrac12}{\nu-\tfrac12}>0
 \qquad (\nu>1/2).
 \label{eq:Sprime}
\end{equation}
At fixed $\nu$,
\begin{equation}
 s=2\sqrt{\nu^2+r^2},
 \qquad
 E=\omega\sqrt{\nu^2+r^2},
 \label{eq:E-nu-r}
\end{equation}
so equal entropy does not imply equal energy on the full Gaussian state space.

The reference dynamics is the standard thermal quantum-optical master
equation \cite{BreuerPetruccione,RivasHuelga,Gorini1976,Lindblad1976},
\begin{equation}
 \dot\rho=-i[H,\rho]
 +\gamma(N+1)\mathcal D[a]\rho
 +\gamma N\mathcal D[a^\dagger]\rho,
 \label{eq:master}
\end{equation}
where
\begin{equation}
 \mathcal D[c]\rho=c\rho c^\dagger-\frac12\{c^\dagger c,\rho\},
 \qquad a=\frac{X+iP}{\sqrt2},
\end{equation}
and
\begin{equation}
 N=\frac{1}{e^{\omega/T_{\rm bath}}-1},
 \quad
 \nu_{\rm th}=N+\frac12
 =\frac12\coth\frac{\omega}{2T_{\rm bath}}.
 \label{eq:nu-th}
\end{equation}
We assume weak system--bath coupling, an initially factorized state, and the
Born--Markov and secular approximations.  In particular,
\begin{equation}
 \gamma\ll\omega,
 \qquad
 \tau_B\ll\gamma^{-1},
 \label{eq:benchmark-hierarchy}
\end{equation}
where $\tau_B$ is the bath correlation time.  Equation~\eqref{eq:master}
preserves Gaussianity and vanishing first moments.  
The benchmark results through Sec.~\ref{sec:operational} assume constant $\omega$, $\gamma$, $T_{\rm bath}$.
Section~\ref{sec:time-dependent-bath} then relaxes the latter two assumptions while keeping $\omega$ fixed.
This deliberate specialization removes any external protocol from the main result.

\section{Exact thermalization of Gaussian second moments}
\label{sec:moments}

Equation~\eqref{eq:master} gives the Lyapunov equation
\begin{equation}
 \dot\sigma=\omega(J\sigma-\sigma J)
 -\gamma(\sigma-\nu_{\rm th}\I),
 \qquad
 J=\begin{pmatrix}0&1\\-1&0\end{pmatrix}.
 \label{eq:lyapunov}
\end{equation}
The Hamiltonian term rotates the covariance ellipse, while the dissipator
drives it toward $\nu_{\rm th}\I$.  Taking the trace gives
\begin{equation}
\dot s=-\gamma(s-2\nu_{\rm th}).
 \label{eq:s-eq}
\end{equation}
The Hamiltonian rotation preserves $\det\sigma$.  Using
$\dd(\det\sigma)/\dd t=\det\sigma\,
\tr(\sigma^{-1}\dot\sigma)$ and
$\tr\sigma^{-1}=s/\nu^2$ for a $2\times2$ matrix, one obtains
\begin{equation}
 \dot\nu=\gamma\left(\frac{\nu_{\rm th}s}{2\nu}-\nu\right).
 \label{eq:nu-eq}
\end{equation}
Finally, differentiating Eq.~\eqref{eq:nu-s-r} gives
\begin{equation}
\frac{\dd}{\dd t}r^2=-2\gamma r^2.
 \label{eq:r-eq}
\end{equation}
Hence
\begin{align}
 s(t)&=2\nu_{\rm th}+[s(0)-2\nu_{\rm th}]e^{-\gamma t},
 \label{eq:s-solution}\\
 r^2(t)&=r^2(0)e^{-2\gamma t},
 \qquad
 r(t)=r(0)e^{-\gamma t}.
 \label{eq:r-solution}
\end{align}
The covariance orientation rotates at angular velocity $\omega$ (the two
anisotropy components rotate at $2\omega$), while its magnitude $r$ decays with
the damping envelope.  A component derivation and the exact matrix solution are
given in Appendix~\ref{app:covariance}.

Equations~\eqref{eq:s-eq} and \eqref{eq:r-eq} show the simple structure behind
the result.  In the variables $(s,r^2)$ the energy-carrying trace and the squared
squeezing relax independently:
\begin{equation}
 s-2\nu_{\rm th}\propto e^{-\gamma t},
 \qquad
 r^2\propto e^{-2\gamma t}.
 \label{eq:two-rates}
\end{equation}
The entropy variable
\begin{equation}
 \nu=\sqrt{\frac{s^2}{4}-r^2}
\end{equation}
mixes these two sectors. 
This mixing prevents $E$ from being a function of $S$ alone.  In the open
system, however, the same covariance dynamics also determines the entropy
rate.  Once the thermal generator is specified, Eq.~\eqref{eq:nu-eq}
therefore provides an additional dynamical relation that allows the
energy-carrying trace $s$ to be eliminated in favor of $\nu$ and $\dot\nu$.
This is the origin of the representation $E=E(S,\dot S)$ derived below.

\section{Exact rate-dependent energetic closure}
\label{sec:closure}

\subsection{Derivation of \texorpdfstring{$E(S,\dot S)$}{E(S,Sdot)}}
\label{sec:derivation}

Solving Eq.~\eqref{eq:nu-eq} for $s$ gives
\begin{equation}
 s=\frac{2\nu}{\nu_{\rm th}}
 \left(\nu+\frac{\dot\nu}{\gamma}\right),
 \qquad \gamma>0.
 \label{eq:s-from-nu}
\end{equation}
Consequently,
\begin{equation}
 E(\nu,\dot\nu)=\frac{\omega\nu}{\nu_{\rm th}}
 \left(\nu+\frac{\dot\nu}{\gamma}\right).
 \label{eq:main-nu}
\end{equation}
For mixed states, $\nu>1/2$, the entropy is invertible and
$\dot S=S'(\nu)\dot\nu$, so
\begin{equation}
 \boxed{
 E(S,\dot S)=\frac{\omega\nu(S)}{\nu_{\rm th}}
 \left[
 \nu(S)+\frac{\dot S}{\gamma S'(\nu(S))}
 \right].
 }
 \label{eq:main-S}
\end{equation}
No expansion in $\gamma$, in the rate, or in the distance from equilibrium has
been made.  Equation~\eqref{eq:main-S} is an exact identity on the mixed-state
interior for Gaussian states evolving under Eq.~\eqref{eq:master}.  At fixed
entropy the energy is affine in the entropy rate,
\begin{equation}
 \left(\frac{\partial E}{\partial\dot S}\right)_{S}
 =\frac{\omega\nu}{\nu_{\rm th}\,\gamma S'(\nu)}>0.
 \label{eq:dEdSdot}
\end{equation}
The pure-state boundary is better described by the regular variable
$q=\nu^2$ introduced below; see Appendix~\ref{app:coordinate-map}.

\subsection{Physical domain and contour plot}
\label{sec:domain}

The whole $(S,\dot S)$ plane is not physically accessible.  From
$r^2\ge0$ and Eq.~\eqref{eq:nu-s-r} one has $s\ge2\nu$, which through
Eq.~\eqref{eq:s-from-nu} becomes
\begin{equation}
 \boxed{
 \dot S\geq\gamma S'(\nu)\left(\nu_{\rm th}-\nu\right).
 }
 \label{eq:physical-domain}
\end{equation}
This is an exact lower boundary, not a compact bound.  Equality is attained
precisely for $r=0$, i.e. by the one-parameter family
\begin{equation}
 \MG=
 \{\sigma=\nu\I\mid \nu\ge1/2\}.
 \label{eq:gibbs-manifold}
\end{equation}
Every point of $\MG$ is a Gibbs state of the same fixed Hamiltonian at some
temperature.  We refer to this one-parameter set as the Gibbs family.  The
bath-selected equilibrium state is the single point
$\nu=\nu_{\rm th}$ on this manifold.

The eliminated variable can be reconstructed explicitly.  Combining
Eqs.~\eqref{eq:s-from-nu} and \eqref{eq:nu-s-r},
\begin{equation}
 \boxed{
 r^2=\nu^2\left[
 \left(\frac{\nu+\dot S/[\gamma S'(\nu)]}{\nu_{\rm th}}\right)^2-1
 \right].
 }
 \label{eq:r-from-Sdot}
\end{equation}
Equation~\eqref{eq:r-from-Sdot} should be understood conditionally on the
thermal generator.  Once $\gamma$ and $\nu_{\rm th}$ are specified, a measured
pair $(S,\dot S)$ determines the value of $r^2$ that is compatible with that generator.  
Indeed, for fixed $S$ and fixed generator, shown in Appendix~\ref{app:coordinate-map},
\begin{equation}
 \frac{\partial\dot S}{\partial (r^2)}
 =
 \frac{\gamma\,\nu_{\rm th}\,S'(\nu)}
 {2\nu\sqrt{\nu^2+r^2}}>0
 \qquad(\gamma>0,\ \nu>1/2),
 \label{eq:monotone}
\end{equation}
so this inference is unique.

This does not define an intrinsic change of coordinates on the Gaussian state
space.  The quantity $r^2$ is determined by the instantaneous state, whereas
$\dot S$ depends additionally on the generator.  Changing the bath coupling
changes $\dot S$ while leaving the instantaneous covariance matrix unchanged.
Equation~\eqref{eq:r-from-Sdot} is therefore a generator-dependent dynamical
inference, not a state-space reparametrization.

The corresponding static description in the $(S,r^2)$ plane is
straightforward.  At fixed entropy, the squeezing raises the energy according
to $E=\omega\sqrt{\nu(S)^2+r^2}$ with $r^2\geq0.$
Thus the familiar Gaussian-state description uses the entropy together with
an explicit internal coordinate, $r^2$.  

\begin{figure*}[t]
\includegraphics[width=.95\linewidth]{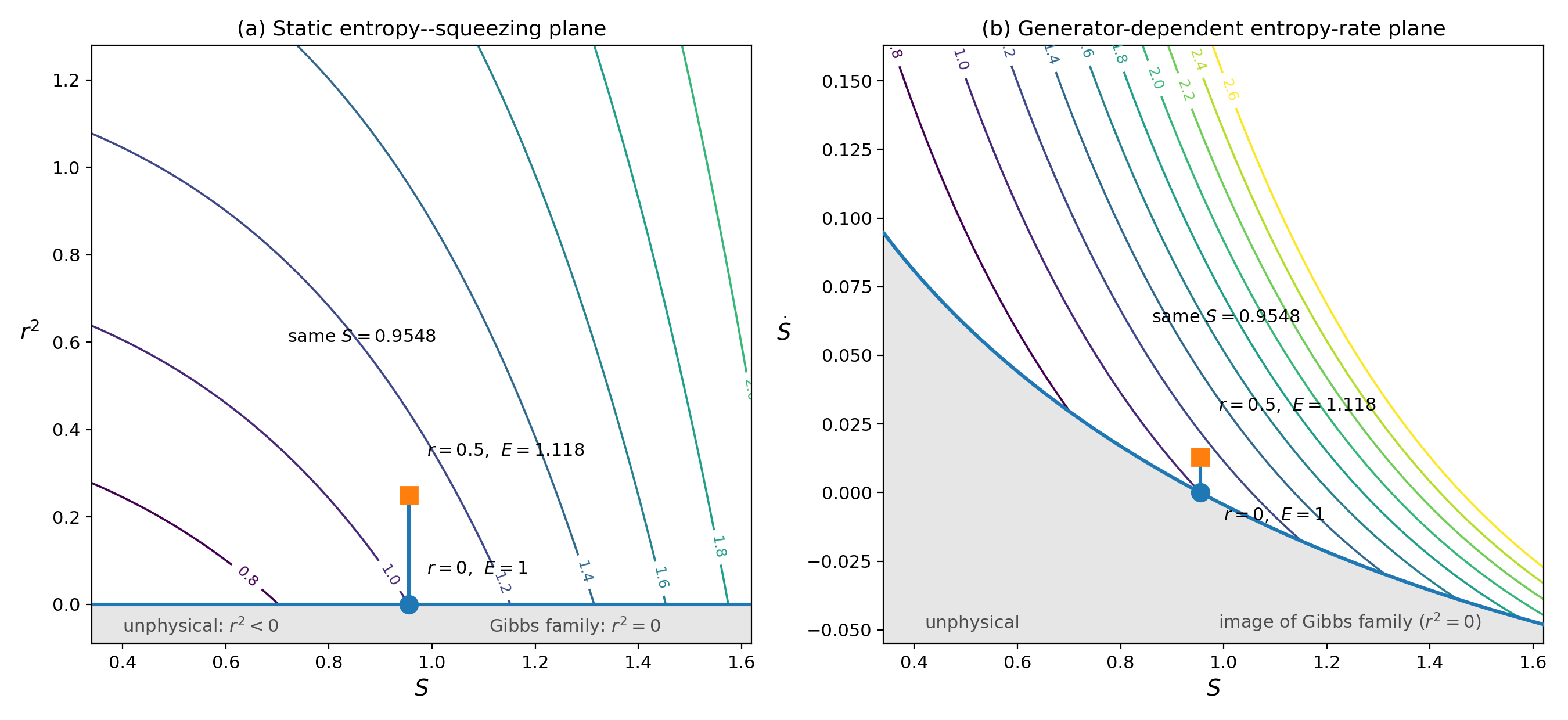}
\caption{
Static state variables and the generator-dependent entropy-rate
representation for $\omega=1$, $\gamma=0.1$, and $\nu_{\rm th}=1$.
(a) In the instantaneous Gaussian-state description, the physical domain is
$r^2\ge0$, with the Gibbs family at $r^2=0$.
(b) For the specified thermal GKLS generator, each Gaussian state acquires an
entropy rate according to Eq.~\eqref{eq:nu-eq}, producing the allowed region
in the $(S,\dot S)$ plane.  Its lower boundary corresponds to unsqueezed
states and is
$\dot S=\gamma S'(\nu)(\nu_{\rm th}-\nu)$.
The marked initial states have the same entropy, $S=0.9548$, but different
squeezing.  Under the specified bath dynamics they consequently have different initial entropy rates and different energies.
}
\label{fig:contours}
\end{figure*}

Figure~\ref{fig:contours} compares these two descriptions directly.
Panel~(a) shows the static $(S,r^2)$ representation, where the physical
boundary is simply $r^2=0$.  
Panel~(b) should not be interpreted as a second coordinate chart on the same
instantaneous state space.  It shows instead the entropy rates assigned to
Gaussian states by the specified thermal generator.  The dependence on
$\gamma$ is essential: the same instantaneous state would occupy a different
point in the $(S,\dot S)$ plane under a different generator.

Under the relaxation map, the straight boundary $r^2=0$ is carried into
\begin{equation}
 \dot S=  \gamma S'(\nu)\bigl(\nu_{\rm th}-\nu\bigr),
\end{equation}
which is the curved lower boundary of the physically accessible region in the
$(S,\dot S)$ plane.
The two marked states in Fig.~\ref{fig:contours} make the correspondence
explicit.  They have the same symplectic eigenvalue, and hence the same
entropy, but different squeezing and therefore different energies.
Panel (a) shows that the states differ in their instantaneous anisotropy, whereas panel (b) shows that the specified thermal generator assigns different entropy rates to those states.

The horizontal line $\dot S =0$ is not an equilibrium manifold; equilibrium is the single intersection of $\dot S=0$ with the Gibbs-family boundary at $ \nu = \nu_{\rm th}$.
The figure also includes a region with $\dot S<0$.
This does not violate the second law: $\dot S$ is the oscillator entropy rate,
not the total entropy production.  Appendix~\ref{app:entropy-production}
makes this distinction explicit using Spohn's inequality.

For the parameters shown, the two marked states have $\nu=1$ with $r=0$ and
$r=0.5$; they share the entropy $S=0.9548$ and correspond respectively to
$\dot S=0$ and $\dot S\simeq1.30\times10^{-2}$.  
The point is that, once the thermal generator is specified, the entropy rate provides the dynamical relation needed to determine the energy without independently measuring the squeezing.

\subsection{Why thermalization is essential}
\label{sec:why-thermalization}

The role of thermalization can be seen most directly by comparing the
open and isolated oscillators.  For the thermal GKLS dynamics,
Eq.~\eqref{eq:nu-eq} gives
\begin{equation}
\dot\nu
=
\gamma
\left(
\frac{\nu_{\rm th}s}{2\nu}-\nu
\right),
\label{eq:nu-rate-thermalization}
\end{equation}
and therefore, for every specified generator with $\gamma>0$, the
energy-carrying variable $s$ can be eliminated algebraically 
to get Eq.~\eqref{eq:s-from-nu}.
This is the dynamical input behind
Eq.~\eqref{eq:main-S}.

It is important that $\dot S$ in this relation is not an additional
instantaneous coordinate of the Gaussian state.  Its value depends on both
the state and the open-system generator.  Using
$s=2\sqrt{\nu^2+r^2}$, for example,
\begin{equation}
\dot S
=
\gamma S'(\nu)
\left[
\frac{\nu_{\rm th}}{\nu}
\sqrt{\nu^2+r^2}
-\nu
\right].
\label{eq:Sdot-generator-dependent}
\end{equation}
Thus the same instantaneous Gaussian state would in general have a different
entropy rate if the damping rate or the bath temperature were changed.
The representation $E(S,\dot S)$ is therefore not a reparametrization of
the instantaneous Gaussian state space.  Rather, it is a relation between
the energy and trajectory-level data generated by a specified thermalizing
dynamics.

The distinction becomes sharp in the isolated limit.  Setting $\gamma=0$
in Eqs.~\eqref{eq:s-eq}--\eqref{eq:r-eq} gives
$\dot s=
\dot r=
\dot\nu=0$
and hence
\begin{equation}
\dot S=0
\end{equation}
for every Gaussian state.  Nevertheless, states with the same $\nu$, and
therefore the same entropy, can have different squeezing $r$ and hence
different energies,
\begin{equation}
E=\omega\sqrt{\nu^2+r^2}.
\end{equation}
Consequently, the pair $(S,\dot S)$ cannot determine the energy of the
isolated oscillator: all equal-entropy states have the same $\dot S=0$
regardless of their energy.

For any nonzero damping rate within the regime of validity of the thermal
master equation, Eq.~\eqref{eq:nu-rate-thermalization} supplies precisely the
additional dynamical relation needed to eliminate $s$ and obtain
Eq.~\eqref{eq:main-S}.  The limit $\gamma\to0^+$ should therefore be
distinguished from the exactly isolated case.  Along a thermalizing
trajectory $\dot\nu=O(\gamma)$, so the ratio $\dot\nu/\gamma$ in
Eq.~\eqref{eq:s-from-nu} may remain finite for arbitrarily
weak but nonzero coupling.  At $\gamma=0$ itself, however, the dynamical
relation disappears and the entropy-rate closure is lost.

In this precise sense, the derivative dependence in
\begin{equation}
E=E(S,\dot S)
\end{equation}
is induced by thermalization.  It does not arise from the speed of an
externally imposed protocol: in the benchmark considered here the Hamiltonian
frequency is constant and no external driving is present.  Instead, the bath
provides a dynamical relation between the instantaneous state and its entropy
rate, and it is this relation that makes the reduced energetic representation
possible.

\subsection{Closed trajectory equation for an entropy-related variable}
\label{sec:entropy-coordinate}

For a specified thermal generator, the trajectory-level closure admits a particularly simple form in terms of $q= \nu^2$.
The variable $q$ is a monotonic reparametrization of the entropy in the
mixed-state interior and remains regular at the pure-state boundary.
Since $\dot q=2\nu\dot\nu$, Eq.~\eqref{eq:main-nu} becomes
\begin{equation}
 E(q,\dot q)
 =
 \frac{\omega}{\nu_{\rm th}}
 \left(q+\frac{\dot q}{2\gamma}\right).
 \label{eq:E-q-qdot}
\end{equation}
At the same time, Eq.~\eqref{eq:s-eq} implies
\begin{equation}
 \dot E=-\gamma(E-\omega\nu_{\rm th}).
 \label{eq:E-relax}
\end{equation}
Eliminating $E$ between Eqs.~\eqref{eq:E-q-qdot} and \eqref{eq:E-relax} gives
\begin{equation}
 \boxed{
 \ddot q+3\gamma\dot q
 +2\gamma^2(q-\nu_{\rm th}^2)=0.
 }
 \label{eq:q-second-order}
\end{equation}
Thus the original first-order reduced dynamics of two Gaussian state variables
has been rewritten as a second-order equation for one entropy-related
coordinate.

The origin of Eq.~\eqref{eq:q-second-order} is transparent once the two
relaxation channels are separated.  Equations \eqref{eq:s-solution} and
\eqref{eq:r-solution} give $s-2\nu_{\rm th}\propto e^{-\gamma t}$ and
$r^2\propto e^{-2\gamma t}$, and $q=s^2/4-r^2$ is quadratic in the first and
linear in the second.  Squaring therefore produces no exponent other than those
already present, so that
\begin{equation}
 q(t)-\nu_{\rm th}^2
 =
 C_1e^{-\gamma t}+C_2e^{-2\gamma t},
 \label{eq:q-general-solution}
\end{equation}
with
\begin{equation}
 C_1=\nu_{\rm th}\left[s(0)-2\nu_{\rm th}\right],
 \quad
 C_2=\tfrac14\left[s(0)-2\nu_{\rm th}\right]^2-r^2(0).
 \label{eq:q-constants}
\end{equation}
Equation~\eqref{eq:q-second-order} is thus nothing but
$(\dd_t+\gamma)(\dd_t+2\gamma)\left[q-\nu_{\rm th}^2\right]=0$.  What is not
automatic is that a quantity built from a square root of a determinant should be
an exact affine combination of the two relaxation modes; this is what makes the
second-order equation linear and closed.  
For a specified thermal generator, admissible initial data
$(q(0),\dot q(0))$ determine a unique trajectory of
Eq.~\eqref{eq:q-second-order}.  These data should not be interpreted as
intrinsic coordinates of the instantaneous Gaussian state: $\dot q(0)$
depends on the generator as well as on that state.  The physical-domain
condition derived in Appendix~\ref{app:coordinate-map} specifies which
initial position--velocity pairs are compatible with Gaussian states under
the chosen generator.

Since $q=q(S)$, Eq.~\eqref{eq:q-second-order} can also be written directly as
an entropy equation,
\begin{equation}
 q'(S)\ddot S
 +q''(S)\dot S^2
 +3\gamma q'(S)\dot S
 +2\gamma^2[q(S)-\nu_{\rm th}^2]
 =0.
 \label{eq:S-second-order}
\end{equation}
In this precise sense, the entropy trajectory can be described by a closed second-order equation once the generator is fixed: $S$ specifies the coordinate and $\dot S$ supplies the second datum needed to determine its trajectory.  
We note, however, that the coordinate in which the reduced dynamics is linear is $q=\nu^2$ and not $S$ itself; Eq.~\eqref{eq:S-second-order} is nonlinear because $q(S)$ is.  
The variable in which the closed trajectory equation becomes linear is therefore $q=\nu^2$, rather than $S$ itself.

There is a second reason for preferring the derivative representation.
The squeezing variable $r^2$ is specific to the single-mode Gaussian
oscillator, and a different open system may require a different internal
variable to complete its instantaneous energetic description.  The
elimination mechanism itself is more general.  Suppose that a state
coordinate $Y$ and an additional internal variable $Z$ determine the
energy according to
\begin{equation}
 E=\Phi(Y,Z),
\end{equation}
while the relaxation dynamics gives
\begin{equation}
 \dot Y=\mathcal R(Y,Z).
\end{equation}
Whenever this relation can be inverted locally for $Z$, the internal
variable may be eliminated,
\begin{equation}
 Z=Z(Y,\dot Y),
 \qquad
 E=\widetilde E(Y,\dot Y).
\end{equation}
For the oscillator studied here, $Y$ may be taken as the entropy-related
coordinate and $Z$ as the squeezing information.  Thus the significance
of the $(S,\dot S)$ representation is not that $\dot S$ is universally
preferable to an internal variable, but that a specified relaxation law can make the internal variable inferable from state-and-rate data.

\subsection{Operational content}
\label{sec:operational}

The generator dependence of $\dot S$ gives
Eq.~\eqref{eq:main-S} direct operational content.  The test does not require
reconstructing the squeezing variable $r^2$.  Instead, the mean energy and
the state purity can be measured independently as functions of time and their
trajectories compared with the relation predicted by the thermal GKLS
generator.

For a one-mode Gaussian state, the purity is
\begin{equation}
 \mathcal P
 :=
 \tr\rho^2
 =
 \frac{1}{2\nu},
 \label{eq:purity}
\end{equation}
so that a direct purity measurement determines $\nu$, and hence the entropy,
without resolving the orientation or anisotropy of the covariance matrix.
The energy is independently given by
\begin{equation}
 E=\omega\left(\bar n+\frac12\right)
   =\frac{\omega s}{2}.
\end{equation}
Thus the test requires scalar energetic and purity information rather than
full covariance tomography.  The price is that the purity must be measured
as a time-resolved quantity.

A particularly useful form of the prediction avoids numerical
differentiation altogether.  Introducing
\begin{equation}
 q:=\nu^2,
\end{equation}
Eq.~\eqref{eq:nu-eq} gives
\begin{equation}
 \dot q+2\gamma q
 =
 \gamma\nu_{\rm th}s
 =
 \frac{2\gamma\nu_{\rm th}}{\omega}E(t).
 \label{eq:operational-q}
\end{equation}
Hence, for a specified thermal generator,
\begin{equation}
 \boxed{
 q(t)
 =
 e^{-2\gamma(t-t_0)}q(t_0)
 +
 \frac{2\gamma\nu_{\rm th}}{\omega}
 \int_{t_0}^{t}
 e^{-2\gamma(t-\tau)}E(\tau)\,\dd\tau .
 }
 \label{eq:operational-integral}
\end{equation}
Since $q=1/(4\mathcal P^2)$, the measured energy trajectory together with the
single initial purity $\mathcal P(t_0)$ predicts the complete purity, and
therefore entropy, trajectory.

In practice, $\gamma$ and $\nu_{\rm th}$ may be determined independently from
the bath temperature and an energy-relaxation calibration.  For a nonstationary energy trace they may also be obtained from a fit to
\begin{equation}
 E(t)
 =
 \omega\nu_{\rm th}
 +
 \left[E(t_0)-\omega\nu_{\rm th}\right]
 e^{-\gamma(t-t_0)}.
\end{equation}
Once these generator parameters and the initial purity are fixed,
Eq.~\eqref{eq:operational-integral} contains no additional dynamical fit
parameters.  The measured purity trajectory therefore provides an independent
test of the thermal Gaussian relaxation law.

The importance of this test is that the energy relaxation alone does not
determine the full dissipative dynamics.  To see this, replace the isotropic
thermal diffusion matrix by
\begin{equation}
 D=\gamma\,\mathrm{diag}(d_1,d_2),
 \qquad
 \bar d=\frac{d_1+d_2}{2}.
\end{equation}
For two reservoirs with the same damping rate and the same trace of $D$,
the trace equation is identical,
\begin{equation}
 \dot s=-\gamma(s-2\bar d),
\end{equation}
and therefore the same initial state produces the same energy trajectory.
The symplectic eigenvalue, however, obeys
\begin{equation}
 \frac{\dd q}{\dd t}
 +2\gamma q
 -\gamma\bar d\,s
 =
 \gamma(d_2-d_1)u,
 \qquad
 u=\frac{\sigma_{11}-\sigma_{22}}{2}.
 \label{eq:aniso-residual}
\end{equation}
Thus reservoirs that are indistinguishable through $E(t)$ can be
distinguished through their purity or entropy trajectories.

For the thermal benchmark one has
$d_1=d_2=\nu_{\rm th}$, so the right-hand side of
Eq.~\eqref{eq:aniso-residual} vanishes identically and
Eq.~\eqref{eq:operational-integral} follows.  An anisotropic diffusion matrix
instead sources the anisotropy component $u$ and produces a nonzero residual.
For generic transient initial conditions this residual contains a damped
component at the orientation frequency $2\omega$; at finite anisotropy it can
also contain a stationary contribution.  The detailed evolution is given in
Appendix~\ref{app:coordinate-map}.

Operationally, Eq.~\eqref{eq:main-S} is therefore not merely a way of
rewriting the internal energy.  Once the thermal generator has been
calibrated, it gives an independent consistency relation between two scalar
time series, energy and purity.  A failure of the measured purity trajectory
to satisfy Eq.~\eqref{eq:operational-integral} signals a failure of the assumed
thermal isotropic relaxation law, with anisotropic diffusion providing one
explicit example.

\subsection{Time-dependent bath temperature}
\label{sec:time-dependent-bath}

We first retain a fixed oscillator frequency but allow the bath temperature and,
if desired, the damping rate to vary in time.  Within a time-local thermal GKLS
description,
\begin{equation}
 \nu_{\rm th}\to\nu_{\rm th}(t)
 =
 \frac12\coth\frac{\omega}{2T_{\rm bath}(t)},
 \qquad
 \gamma\to\gamma(t),
\end{equation}
and Eq.~\eqref{eq:lyapunov} retains the same local form.  The instantaneous bath
target is
\begin{equation}
 \sigma_{\rm target}(t)=\nu_{\rm th}(t)\I.
\end{equation}
Changing $T_{\rm bath}(t)$ moves the target along the same fixed Gibbs family
$\MG$; it does not change that family itself.

The exact local equations are
\begin{align}
 \dot s&=-\gamma(t)[s-2\nu_{\rm th}(t)],
 \label{eq:s-timebath}\\
 \dot\nu&=\gamma(t)
 \left[\frac{\nu_{\rm th}(t)s}{2\nu}-\nu\right],
 \label{eq:nu-timebath}\\
 \frac{\dd}{\dd t}r^2&=-2\gamma(t)r^2.
 \label{eq:r-timebath}
\end{align}
Hence
\begin{equation}
 r^2(t)=r^2(t_0)
 \exp\left[-2\int_{t_0}^{t}\gamma(u)\,\dd u\right].
 \label{eq:r-timebath-solution}
\end{equation}
If the state starts unsqueezed, it remains unsqueezed.  Bath-temperature
driving at fixed $\omega$ therefore produces motion along the Gibbs family,
not transverse squeezing.

Whenever $\gamma(t)>0$, the same elimination can be performed pointwise:
\begin{equation}
 \boxed{
 E(S,\dot S;t)=
 \frac{\omega\nu(S)}{\nu_{\rm th}(t)}
 \left[
 \nu(S)+
 \frac{\dot S}{\gamma(t)S'(\nu(S))}
 \right].
 }
 \label{eq:main-S-timebath}
\end{equation}
No explicit $\dot T_{\rm bath}$ term appears.  The time-dependent bath enters
through the instantaneous target and relaxation rate.  The explicit
$t$-dependence of Eq.~\eqref{eq:main-S-timebath} is of the same character as the
control-parameter dependence $\lambda$ in the equilibrium form $E=E(S,\lambda)$:
$\nu_{\rm th}(t)$ and $\gamma(t)$ are externally prescribed parameters, not
additional state variables.

\section{Driven oscillator: generation and relaxation of off-Gibbs structure}
\label{sec:driven}

The fixed-frequency analysis establishes the exact entropy-rate representation
of the internal energy under standard thermal relaxation.  It does not,
however, provide a mechanism for generating the off-Gibbs Gaussian structure
from an initially Gibbs state.  Indeed, at fixed oscillator frequency,
Eq.~\eqref{eq:r-eq} gives $ \frac{d}{dt}r^2=-2\gamma r^2,$
so that
\begin{equation}
 r(t_0)=0
 \quad\Longrightarrow\quad
 r(t)=0 .
 \label{eq:no-squeezing-fixed}
\end{equation}
The same conclusion holds when only the bath temperature changes:
the thermal target moves along the fixed Gibbs family, but no transverse
Gaussian deformation is generated.  Thus thermal relaxation can remove
pre-existing squeezing, but at fixed $\omega$ it cannot create it for the isotropic thermal generator considered here.

A different ingredient is required if the off-Gibbs structure is to arise
dynamically.  We now show that frequency modulation supplies such a mechanism.
Consider
\begin{equation}
 H(t)=\frac12\left[p^2+\omega^2(t)x^2\right].
 \label{eq:H-driven}
\end{equation}
The driven construction has a different logical status from the
fixed-frequency benchmark.  Equation~\eqref{eq:main-S} follows exactly from
the standard thermal GKLS equation, whereas the construction below is a local
Gaussian continuation whose validity requires additional assumptions on the
bath and driving timescales.  Its purpose is to identify how Hamiltonian
driving can generate the Gaussian structure on which thermal relaxation
subsequently acts.  Technical details are collected in
Appendix~\ref{app:driven-details}.

\subsection{Invariant description of the driven Gaussian state}

Write the zero-mean Gaussian state as
\begin{equation}
 \rho(t)=Z^{-1}\exp[-\mathcal I(t)/T_0],
 \label{eq:rho-driven-main}
\end{equation}
where $T_0$ is a fixed reference scale and
\begin{equation}
 \mathcal I(t)
 =
 \frac12
 \left[
 g_-(t)p^2
 +g_0(t)\{x,p\}
 +g_+(t)x^2
 \right].
 \label{eq:quadratic-kernel}
\end{equation}
For isolated evolution the quadratic kernel satisfies
\begin{equation}
 D_t\mathcal I
 \equiv
 \partial_t\mathcal I+i[H(t),\mathcal I]
 =0 .
 \label{eq:closed-invariant}
\end{equation}
The symplectic kernel scale
\begin{equation}
 \omega_{\mathcal I}
 =
 \sqrt{g_-g_+-g_0^2}
 \label{eq:invariant-frequency}
\end{equation}
is therefore exactly constant under the unitary dynamics, even when
$\omega(t)$ varies
\cite{Ermakov1880,Lewis1969,LewisLeach1982,Ray1982,Kim:1996vm}.
The drive changes the shape and orientation of the quadratic kernel, but not
this invariant scale.

It is also useful to define
\begin{equation}
 \omega_{\rm eff}
 =
 \frac{g_++\omega^2 g_-}{2\omega_{\mathcal I}},
 \label{eq:omega-effective}
\end{equation}
which determines the physical energy.  The exact Gaussian dictionary is
\begin{equation}
 \nu
 =
 \frac12\coth\frac{\omega_{\mathcal I}}{2T_0},
 \qquad
 E=\nu\,\omega_{\rm eff}.
 \label{eq:driven-dictionary-main}
\end{equation}
Thus $\omega_{\mathcal I}$ determines the mixedness of the Gaussian state,
whereas $\omega_{\rm eff}$ measures its energetic deformation relative to
the instantaneous oscillator Hamiltonian.  On the instantaneous Gibbs family,
\begin{equation}
 \mathcal I=\lambda H(t),
\end{equation}
one has
\begin{equation}
 \omega_{\rm eff}=\omega(t),
\end{equation}
while nonadiabatic deformation generically produces
$\omega_{\rm eff}\neq\omega$~\cite{Kim:2021wyp}.

\subsection{Local relaxation toward a moving Gibbs target}

To describe weak thermal relaxation in the driven problem, we adopt the local
Gaussian continuation
\begin{equation}
 \boxed{
 D_t\mathcal I
 =
 -\alpha(t)
 \left[
 \mathcal I-g(t)H(t)
 \right],
 }
 \qquad
 g(t)=\frac{T_0}{T_{\rm bath}(t)}.
 \label{eq:local-relaxation-law}
\end{equation}
Extensions of invariant methods to open systems have been considered
previously~\cite{Chen2015}; here Eq.~\eqref{eq:local-relaxation-law} is used
specifically as a local relaxation model anchored to the fixed-frequency
thermal problem.  It is not assumed to be a universal microscopic GKLS
generator for arbitrary driving.

The origin of the off-Gibbs structure is most transparent if we define the
deviation from the instantaneous thermal target,
\begin{equation}
 \delta\mathcal I
 :=
 \mathcal I-gH .
 \label{eq:deltaI-driven}
\end{equation}
Equation~\eqref{eq:local-relaxation-law} then gives
\begin{equation}
 \boxed{
 D_t\delta\mathcal I+\alpha\delta\mathcal I
 =
 -\dot g\,H-g\dot H .
 }
 \label{eq:moving-target-main}
\end{equation}
The two source terms have different roles.  At fixed $\omega$, the term
$\dot g\,H$ moves the target only along the Gibbs family.  Frequency
modulation is qualitatively different because
\begin{equation}
 \dot H
 =
 \omega\dot\omega\,x^2 .
\end{equation}
In terms of the instantaneous ladder operator
\begin{equation}
 a_t
 =
 \sqrt{\frac{\omega}{2}}\,x
 +
 \frac{i}{\sqrt{2\omega}}\,p ,
\end{equation}
this becomes
\begin{equation}
 \dot H
 =
 \frac{\dot\omega}{2}
 \left(
 a_t^2+a_t^{\dagger 2}
 +2a_t^\dagger a_t+1
 \right).
 \label{eq:Hdot-pair-main}
\end{equation}
The terms $a_t^2$ and $a_t^{\dagger2}$ are transverse to the instantaneous
Gibbs family and generate pair coherence, or equivalently squeezing.
Consequently, even if the state initially coincides with the instantaneous
thermal target,
\begin{equation}
 \delta\mathcal I(t_0)=0,
\end{equation}
a nonzero $\dot\omega(t_0)$ generically drives it away from that family.

The physical division of labor is therefore
\begin{align*}
&\text{frequency driving}\\
 &\longrightarrow
 \text{off-Gibbs Gaussian deformation, thermal relaxation}\\
& \longrightarrow
 \text{decay of that deformation}.
 \label{eq:drive-relax-roles}
\end{align*}
This is the mechanism absent from the fixed-frequency benchmark.

\subsection{Relaxation of the invariant scale and energetic representation}

The local relaxation law also determines the evolution of the invariant
kernel scale.  The component equations following from
Eq.~\eqref{eq:local-relaxation-law} imply
\begin{equation}
 \boxed{
 \dot\omega_{\mathcal I}
 =
 \alpha
 \left(
 g\omega_{\rm eff}-\omega_{\mathcal I}
 \right).
 }
 \label{eq:omegaI-relaxation}
\end{equation}
This equation cleanly separates the unitary and dissipative roles.
For $\alpha=0$,
\begin{equation}
 \dot\omega_{\mathcal I}=0
\end{equation}
even under nonadiabatic frequency driving.  The drive can therefore generate
squeezing while leaving the entropy unchanged in the isolated system.
Only the coupling to the bath makes $\omega_{\mathcal I}$, and hence the
entropy, evolve.

Solving Eq.~\eqref{eq:omegaI-relaxation} for the energetic scale gives
\begin{equation}
 \omega_{\rm eff}
 =
 \frac{1}{g}
 \left(
 \omega_{\mathcal I}
 +\frac{\dot\omega_{\mathcal I}}{\alpha}
 \right).
 \label{eq:omegaeff-driven-eliminated}
\end{equation}
For $\alpha(t)>0$, solving
Eq.~\eqref{eq:omegaI-relaxation} for the energetic scale gives
\begin{equation}
 \boxed{
 E
 =
 \frac{1}{2g}
 \left(
 \omega_{\mathcal I}
 +\frac{\dot\omega_{\mathcal I}}{\alpha}
 \right)
 \coth\frac{\omega_{\mathcal I}}{2T_0}.
 }
 \label{eq:driven-energy-reduced}
\end{equation}
Since the Gaussian entropy is determined by $\omega_{\mathcal I}$,
Eq.~\eqref{eq:driven-energy-reduced} is the driven analogue of the
generator-conditioned entropy-rate representation found in the
fixed-frequency problem.

This statement should not be interpreted as an intrinsic change of coordinates
on Gaussian state space.  The quantity $\dot\omega_{\mathcal I}$, like
$\dot S$, depends on the specified dissipative dynamics.  Equation
~\eqref{eq:driven-energy-reduced} is therefore a trajectory-level energetic
relation conditioned on the local relaxation model
\eqref{eq:local-relaxation-law}.

\subsection{Relation to the fixed-frequency benchmark}

The fixed-frequency and driven relations have different levels of control.
At fixed $\omega$, Eq.~\eqref{eq:main-S} is the appropriate result because it
follows directly from the standard thermal master equation
\eqref{eq:master}.  Equation~\eqref{eq:driven-energy-reduced}, by contrast,
follows from the additional modeling assumption
\eqref{eq:local-relaxation-law}.

The two become quantitatively equivalent at fixed frequency only when the
rate $\alpha$ is chosen so that the invariant-kernel evolution reproduces the
GKLS entropy trajectory.  The corresponding matching condition is derived in
Appendix~\ref{app:driven-details}.  In particular, $\alpha$ should not in
general be identified with the benchmark damping rate $\gamma$ away from the
near-target regime.

The two calculations therefore answer complementary questions.  The
fixed-frequency benchmark establishes that thermalization gives an exact
relation between the internal energy and the entropy rate.  The driven
construction shows how the nonequilibrium Gaussian structure entering that
relation can itself be generated dynamically:
\begin{align*}
 &\text{frequency modulation} \\
 &\;\longrightarrow\;
 \text{squeezing} \\
& \;\longrightarrow\;
 \text{thermal relaxation}\\
 &\;\longrightarrow\;
 \text{nontrivial entropy dynamics}.
 \label{eq:driven-physical-chain}
\end{align*}
A driven but isolated oscillator realizes the first two steps but has
$\dot S=0$; an undriven thermal oscillator realizes relaxation but cannot
generate squeezing from a Gibbs initial state.  Both ingredients are required
for the full dynamical process.

It is useful to distinguish two independently controlled facts from their
combination in the driven open system.  First, nonadiabatic unitary frequency
modulation can generically generate off-Gibbs Gaussian structure while leaving
the von Neumann entropy unchanged; this follows from the isolated dynamics and
does not rely on Eq.~\eqref{eq:local-relaxation-law}.  
Second, in the exactly controlled fixed-frequency thermal GKLS benchmark, off-Gibbs structure affects the entropy rate through Eq.~\eqref{eq:nu-eq}.  Combining these two mechanisms into a single driven open-system trajectory, and obtaining the quantitative relation~\eqref{eq:driven-energy-reduced}, requires the local Gaussian continuation~\eqref{eq:local-relaxation-law}.  
The qualitative division of labor---unitary driving generates off-Gibbs
structure, while thermal relaxation makes its subsequent evolution contribute
to entropy change---is established by the two independently controlled limits;
only their quantitative synthesis in a driven open system depends on the local continuation.

\subsection{Regime of validity}

The local construction requires the bath to lose memory before the resonant
data entering the system--bath coupling vary appreciably.  Let
\begin{equation}
 \tau_{\mathcal I}
 =
 \left|
 \frac{\omega_{\mathcal I}}
 {\dot\omega_{\mathcal I}}
 \right|,
 \qquad
 \tau_A
 =
 \left|
 \frac{A}{\dot A}
 \right|,
\end{equation}
where $A(t)$ denotes the matrix-element prefactor in the decomposition of the
system coupling operator in the invariant basis.  A compact Markov hierarchy is
\begin{equation}
 \boxed{
 \tau_B
 \ll
 \min
 \left(
 \alpha^{-1},
 \tau_{\mathcal I},
 \tau_A
 \right).
 }
 \label{eq:driven-markov-summary}
\end{equation}
If a secular coarse-graining time $\Delta t$ is introduced, one further
requires
\begin{equation}
 \boxed{
 \max
 \left(
 \tau_B,
 \omega_{\mathcal I}^{-1}
 \right)
 \ll
 \Delta t
 \ll
 \min
 \left(
 \alpha^{-1},
 \tau_{\mathcal I},
 \tau_A
 \right).
 }
 \label{eq:driven-secular-summary}
\end{equation}

These conditions delimit the status of the driven construction.  The
fixed-frequency entropy-rate relation remains the exactly controlled
benchmark; the driven model supplies a local realization of how frequency
modulation generates transverse Gaussian structure and how subsequent
thermalization converts that structure into nontrivial entropy dynamics.
The full covariance--kernel dictionary, component evolution equations,
matching to the fixed-frequency GKLS benchmark, and derivation of the
timescale conditions are given in Appendix~\ref{app:driven-details}.

\section{Discussion}
\label{sec:discussion}

The analysis above separates two logically distinct ingredients of the
problem.  The first is thermal relaxation itself.  For the standard
fixed-frequency thermal GKLS oscillator, the covariance equations imply the
exact mixed-state relation
\begin{equation}
 E=E(S,\dot S),
\end{equation}
with the explicit form given in Eq.~\eqref{eq:main-S}.  No external driving is
required for this result.  The entropy rate is not an instantaneous state
coordinate: it is determined jointly by the Gaussian state and the specified
open-system generator.  The entropy-rate representation is therefore a
trajectory-level energetic relation conditioned on thermalizing dynamics, not
a reparametrization of the instantaneous Gaussian state space.

The isolated limit shows why the bath is essential to this representation.
When $\gamma=0$, the symplectic eigenvalue and hence the entropy are constants
of motion, so that $\dot S=0$ for every Gaussian state.  Equal-entropy states
can nevertheless have different energies because they may carry different
squeezing.  The entropy rate therefore contains no additional energetic
information in the isolated oscillator.  For nonzero thermal coupling,
Eq.~\eqref{eq:nu-eq} supplies the additional dynamical relation that allows the
energy-carrying second moment to be eliminated in favor of $S$ and $\dot S$.
Within the fixed-frequency benchmark, this is the precise sense in which the
rate dependence is induced by thermalization.

Thermalization, however, does not by itself generate the off-Gibbs structure
that makes this relation nontrivial.  At fixed oscillator frequency,
\begin{equation}
 \frac{d}{dt}r^2=-2\gamma r^2,
\end{equation}
so an initially unsqueezed state remains unsqueezed.  The same remains true
when only the bath temperature varies: the bath target moves along the fixed
Gibbs family, while no transverse squeezing is produced.  Thus the
fixed-frequency problem establishes the exact energetic relation and describes
the relaxation of pre-existing nonequilibrium Gaussian structure, but it does
not provide a mechanism for creating that structure from an initially Gibbs
state.

This is the physical role of the driven construction.  When the oscillator
frequency varies, the instantaneous Gibbs family itself moves.  The tracking
equation
\begin{equation}
 D_t\delta\mathcal I+\alpha\delta\mathcal I
 =
 -\dot g\,H-g\dot H
\end{equation}
shows that the two sources have different geometric effects.  The term
$\dot g\,H$ moves the target tangentially along the Gibbs family, whereas
$\dot H$ contains pair terms in the instantaneous oscillator basis and
generically drives the state transversely away from that family.  Frequency
modulation therefore provides a physical mechanism for generating squeezing,
while the bath subsequently relaxes the generated deformation.

The roles of driving and thermalization are consequently complementary.
A driven but isolated oscillator can develop squeezing and an energetic
departure from the instantaneous Gibbs family, but its von Neumann entropy
remains constant and no entropy-rate closure follows.  Conversely, an
undriven thermal oscillator supplies the exact relation $E(S,\dot S)$ but
cannot generate squeezing from an initially Gibbs state.  In the driven open
oscillator the two mechanisms operate together:
\begin{align}
& \text{frequency driving} \nonumber \\
& \;\longrightarrow\; 
 \text{off-Gibbs Gaussian deformation} \nonumber \\
& \;\longrightarrow\; 
 \text{thermal relaxation and nontrivial }\dot S . \nonumber
\end{align}
This combined process is the dynamical realization of the entropy-rate
energetic structure identified by the fixed-frequency benchmark.

This separation also clarifies the relation to finite-time thermodynamics.
The appearance of $\dot S$ in the energetic representation is not caused by
the speed of the driving protocol: Eq.~\eqref{eq:main-S} already holds for a
time-independent Hamiltonian.  External driving has a different role here.
It generates the transverse Gaussian structure on which the bath subsequently
acts.  The derivative dependence is supplied by the relaxation law, whereas
the drive supplies a mechanism for producing the nonequilibrium structure
whose relaxation makes that dependence physically active.

The two parts of the analysis have different levels of generality.  The
fixed-frequency result follows exactly from the standard weak-coupling
Born--Markov--secular master equation.  The driven construction is instead a
local Gaussian continuation based on
Eq.~\eqref{eq:local-relaxation-law}.  Its purpose is not to claim a universal
microscopic GKLS equation for an arbitrarily driven bath, but to exhibit a
controlled realization in which the unitary drive generates transverse
Gaussian structure while the invariant scale remains suitable for a local
relaxation description.  Its validity therefore requires the hierarchy
summarized in Eqs.~\eqref{eq:driven-markov-summary} and
\eqref{eq:driven-secular-summary}, and the rate $\alpha$ should not in general
be identified with the fixed-frequency rate $\gamma$ away from the matching
regime.

Because the entropy rate is fixed jointly by the state
and the generator, the relation is falsifiable without any
reconstruction of the squeezing variable.  The protocol
is given in Sec.~\ref{sec:operational}: two scalar time
series, the purity and the mean energy, are measured
independently, and Eq.~\eqref{eq:main-S} asserts that they
are not independent of one another.  What such a test
probes is accordingly the assumed structure of the
dissipator rather than any new energetic effect of
squeezing.  An anisotropic reservoir illustrates the
distinction, since it can reproduce the energy trajectory
exactly while failing the entropy-rate relation.

One point of interpretation should be kept separate from
this.  The physically accessible region of
Fig.~\ref{fig:contours} contains states with $\dot S<0$, so
the oscillator entropy may decrease along a thermalizing
trajectory.  Only the total entropy production, which also
counts the flow to the bath, is constrained,
\begin{equation}
 \dot\Sigma
 =
 \dot S-\frac{\dot Q}{T_{\rm bath}}
 \ge0 ,
 \label{eq:discussion-spohn}
\end{equation}
established for the same GKLS generator in
Appendix~\ref{app:entropy-production}.

Nothing in this construction changes the usual definition
$E=\tr(\rho H)$ as an instantaneous state function on the full Gaussian state
space.  The new statement concerns a reduced, generator-conditioned
representation of that energy.  The fixed-frequency oscillator shows that
thermal relaxation permits this representation to close exactly on
$(S,\dot S)$, while the driven oscillator shows how the off-Gibbs Gaussian
structure entering that relation can be generated dynamically rather than
inserted through the initial condition.

The same logic suggests extensions to multimode Gaussian systems, but with an
important qualification.  If independent modes each obey an appropriate
thermal Gaussian relaxation law, one may obtain mode-resolved relations
\begin{equation}
 E_{\bm k}=E_{\bm k}(S_{\bm k},\dot S_{\bm k}).
\end{equation}
This does not imply a closed scalar relation
$E_{\rm tot}=E(S_{\rm tot},\dot S_{\rm tot})$, because the distribution of
entropy and entropy rate among the modes generally carries additional
information.  Determining when an analogous generator-dependent elimination
closes in multimode or non-Gaussian systems remains an open problem.

\begin{acknowledgments}
This was supported by Korea National University of Transportation Industry-Academy Cooperation Foundation In 2023.
\end{acknowledgments}

\appendix
\section{Component covariance dynamics}
\label{app:covariance}

This Appendix derives the invariant equations used in
Sec.~\ref{sec:moments} directly from the covariance components and makes the
decoupling of the covariance orientation explicit.

Write
\begin{equation}
 \sigma=
 \begin{pmatrix}
 a & c\\
 c & b
 \end{pmatrix}.
 \label{eq:app-sigma-components}
\end{equation}
Substituting this into Eq.~\eqref{eq:lyapunov} gives
\begin{align}
 \dot a&=2\omega c-\gamma(a-\nu_{\rm th}),
 \label{eq:app-adot}\\
 \dot b&=-2\omega c-\gamma(b-\nu_{\rm th}),
 \label{eq:app-bdot}\\
 \dot c&=\omega(b-a)-\gamma c.
 \label{eq:app-cdot}
\end{align}
The trace equation follows immediately:
\begin{equation}
 \dot s=\dot a+\dot b=-\gamma(s-2\nu_{\rm th}).
\end{equation}

Introduce the two anisotropy components $ u=(a-b)/2$ and $
 v=c.$
They obey
\begin{align}
 \dot u&=2\omega v-\gamma u,
 \label{eq:app-udot}\\
 \dot v&=-2\omega u-\gamma v.
 \label{eq:app-vdot}
\end{align}
Thus the anisotropy vector $(u,v)$ rotates at angular frequency $2\omega$ while
its magnitude decays with envelope $e^{-\gamma t}$.  Since 
$ r^2=u^2+v^2,$ one obtains
\begin{equation}
 \frac{\dd}{\dd t}r^2
 =
 2u\dot u+2v\dot v
 =
 -2\gamma r^2.
\end{equation}
This also shows why the orientation is irrelevant for the energetic closure:
the phase of $u+iv$ does not enter either $s$ or $r^2$.

The full covariance solution can be written compactly as
\begin{equation}
 \boxed{
 \sigma(t)
 =
 \nu_{\rm th}\I
 +
 e^{-\gamma t}
 R(t)\,[\sigma(0)-\nu_{\rm th}\I]\,R^{T}(t),
 }
 \label{eq:app-full-sigma}
\end{equation}
where
\begin{equation}
 R(t)=e^{\omega Jt}
 =
 \begin{pmatrix}
 \cos\omega t & \sin\omega t\\
 -\sin\omega t & \cos\omega t
 \end{pmatrix}.
\end{equation}
Equation~\eqref{eq:app-full-sigma} displays the isotropic relaxation and the
Hamiltonian rotation separately.

Finally, since $\nu^2=\det\sigma$,
\begin{align}
 \frac{\dd}{\dd t}\nu^2
 &=
 \det\sigma\,\tr(\sigma^{-1}\dot\sigma)
 \nonumber\\
 &=
 -2\gamma\nu^2+\gamma\nu_{\rm th}s,
\end{align}
where the Hamiltonian commutator contributes zero.  Therefore
\begin{equation}
 \dot\nu
 =
 \gamma\left(\frac{\nu_{\rm th}s}{2\nu}-\nu\right),
\end{equation}
which is Eq.~\eqref{eq:nu-eq}.

\section{Generator-dependent entropy-rate relation, admissible domain,
and dissipator test}
\label{app:coordinate-map}

This Appendix collects three technical results used in
Sec.~\ref{sec:closure}.  First, it characterizes the relation between the
instantaneous squeezing and the entropy rate for a specified thermal GKLS
generator.  Second, it gives the admissible state-and-rate domain and discusses
the pure-state boundary.  Third, it derives the scalar consistency relation
used in Sec.~\ref{sec:operational} and shows explicitly how it fails for an
anisotropic dissipator.

The relation between $r^2$ and $\dot S$ should throughout be understood
conditionally on the generator parameters $\gamma$ and $\nu_{\rm th}$.  It is
not an intrinsic coordinate transformation on Gaussian state space.

\subsection*{Conditional relation and admissible domain}

For fixed $\nu$, Eq.~\eqref{eq:E-nu-r} gives
\begin{equation}
 s=2\sqrt{\nu^2+r^2}.
\end{equation}
Substitution into Eq.~\eqref{eq:nu-eq} yields
\begin{equation}
 \dot S
 =
 \gamma S'(\nu)
 \left[
 \frac{\nu_{\rm th}}{\nu}\sqrt{\nu^2+r^2}-\nu
 \right].
 \label{eq:app-Sdot-r2}
\end{equation}
For a specified thermal generator with $\gamma>0$ and for $\nu>1/2$,
\begin{equation}
 \frac{\partial\dot S}{\partial(r^2)}
 =
 \frac{\gamma\nu_{\rm th}S'(\nu)}
 {2\nu\sqrt{\nu^2+r^2}}
 >0.
 \label{eq:app-jacobian}
\end{equation}
Thus, at fixed entropy and for fixed generator parameters, the entropy rate
determines uniquely the value of $r^2$ compatible with that dynamics.  Solving
Eq.~\eqref{eq:app-Sdot-r2} for $r^2$ reproduces
Eq.~\eqref{eq:r-from-Sdot}.  This is a generator-dependent dynamical inference,
not an identification of $r^2$ and $\dot S$ as intrinsic state coordinates.

The variable
\begin{equation}
 q:=\nu^2
\end{equation}
provides a convenient description of the admissible state-and-rate data.
From Eq.~\eqref{eq:E-q-qdot},
\begin{equation}
 s
 =
 \frac{2}{\nu_{\rm th}}
 \left(
 q+\frac{\dot q}{2\gamma}
 \right),
 \label{eq:app-s-q}
\end{equation}
and therefore
\begin{equation}
 r^2
 =
 \frac{1}{\nu_{\rm th}^2}
 \left(
 q+\frac{\dot q}{2\gamma}
 \right)^2-q.
 \label{eq:app-r2-q}
\end{equation}
A physical Gaussian covariance must satisfy the uncertainty condition
\begin{equation}
 q\ge\frac14
 \label{eq:app-q-uncertainty}
\end{equation}
together with $r^2\ge0$.  Since $s>0$, the physical branch of the latter
condition is
\begin{equation}
 \boxed{
 \dot q
 \ge
 2\gamma
 \left(
 \nu_{\rm th}\sqrt q-q
 \right).
 }
 \label{eq:app-q-domain}
\end{equation}
Equality corresponds to $r=0$.  Equation~\eqref{eq:q-second-order} therefore
defines a closed trajectory equation only together with the admissibility
conditions
\eqref{eq:app-q-uncertainty}--\eqref{eq:app-q-domain};
arbitrary mathematical initial data $(q,\dot q)$ need not correspond to a
physical Gaussian state evolving under the specified generator.

For completeness, the constants in Eq.~\eqref{eq:q-general-solution} are fixed
by $q_0=q(0)$ and $\dot q_0=\dot q(0)$ as
\begin{align}
 C_1
 &=
 2(q_0-\nu_{\rm th}^2)
 +\frac{\dot q_0}{\gamma},
 \\
 C_2
 &=
 -(q_0-\nu_{\rm th}^2)
 -\frac{\dot q_0}{\gamma}.
\end{align}

\subsection{Pure-state boundary}

The entropy representation requires one additional qualification.  At the
pure-state boundary $\nu=1/2$,
\begin{equation}
 S'(\nu)
 =
 \ln\frac{\nu+\tfrac12}{\nu-\tfrac12}
 \longrightarrow\infty .
 \label{eq:app-Sprime-pure}
\end{equation}
Consequently, formulas written directly in terms of $(S,\dot S)$ become
singular at this boundary for generic thermalizing trajectories.  This is why
Eq.~\eqref{eq:main-S} was stated for mixed states.

The variable $q=\nu^2$, by contrast, remains regular at $q=1/4$, and
Eq.~\eqref{eq:E-q-qdot} continues to provide the appropriate state-and-rate
representation there.  The zero-temperature vacuum fixed point is exceptional:
there $\nu=1/2$ and $\dot\nu=0$, and the corresponding expressions should be
understood through the limiting fixed trajectory.

\subsection{Failure for an anisotropic dissipator}
Using $q=\nu^2$, Eq.~\eqref{eq:nu-eq} gives Eq.~\eqref{eq:operational-q} and its integrated solution~\eqref{eq:operational-integral} of the main
text.  Thus, once the thermal generator is specified, an energy trajectory
and a single initial value of $q$ determine the symplectic-eigenvalue
trajectory predicted by the model.

The special role of the isotropic thermal diffusion matrix can be seen by
replacing the diffusion term in Eq.~\eqref{eq:lyapunov} by a general symmetric
matrix $D$.  The covariance equation then takes the form
\begin{equation}
 \dot\sigma
 =
 \omega(J\sigma-\sigma J)
 -\gamma\sigma+D .
 \label{eq:app-general-lyapunov}
\end{equation}
Using $
 \frac{\dd}{\dd t}\det\sigma
 =
 \det\sigma\,
 \tr(\sigma^{-1}\dot\sigma),$
together with
$\det\sigma\,\sigma^{-1}=\operatorname{adj}\sigma$ for a $2\times2$ matrix,
one obtains
\begin{equation}
 \frac{\dd q}{\dd t}
 =
 -2\gamma q
 +
 \tr\!\left(\operatorname{adj}\sigma\,D\right).
 \label{eq:app-q-general}
\end{equation}
The Hamiltonian rotation again drops out of the determinant.

For the thermal isotropic choice $ D=\gamma\nu_{\rm th}\I,$
Eq.~\eqref{eq:app-q-general} reduces to Eq.~\eqref{eq:operational-q}.
Now consider instead
\begin{equation}
 D
 =
 \gamma\,\mathrm{diag}(d_1,d_2),
 \qquad
 \bar d:=\frac{d_1+d_2}{2}.
 \label{eq:app-aniso-D}
\end{equation}
We restrict $d_1> 0$ and $d_2>0$ to values compatible with complete positivity
of the Gaussian Markovian dynamics; in the present normalization this requires $d_1d_2\ge 1/4$.
Writing $\sigma_{11}=\frac{s}{2}+u,$ $\sigma_{22}=\frac{s}{2}-u,$ and 
$ v:=\sigma_{12}$ gives Eq.~\eqref{eq:aniso-residual} of the main text.

The same anisotropic diffusion also modifies the evolution of the orientation
variables themselves.  The covariance components obey
\begin{align}
 \dot\sigma_{11}
 &=
 2\omega\sigma_{12}
 -\gamma\sigma_{11}
 +\gamma d_1,
 \\
 \dot\sigma_{22}
 &=
 -2\omega\sigma_{12}
 -\gamma\sigma_{22}
 +\gamma d_2,
 \\
 \dot\sigma_{12}
 &=
 \omega(\sigma_{22}-\sigma_{11})
 -\gamma\sigma_{12},
\end{align}
and hence
\begin{align}
 \boxed{
 \dot u
 =
 2\omega v-\gamma u
 +\frac{\gamma}{2}(d_1-d_2),
 }
 \label{eq:app-aniso-udot}
 \\
 \boxed{
 \dot v
 =
 -2\omega u-\gamma v .
 }
 \label{eq:app-aniso-vdot}
\end{align}
Unlike the isotropic equations
\eqref{eq:app-udot}--\eqref{eq:app-vdot},
Eq.~\eqref{eq:app-aniso-udot} contains a constant source.  The anisotropy
therefore relaxes not to zero but to
\begin{align}
 u_{\rm ss}
 &=
 \frac{\gamma^2(d_1-d_2)}
 {2(4\omega^2+\gamma^2)},
 \label{eq:app-aniso-uss}
 \\
 v_{\rm ss}
 &=
 -\frac{\gamma\omega(d_1-d_2)}
 {4\omega^2+\gamma^2}.
 \label{eq:app-aniso-vss}
\end{align}
Writing $\Delta t=t-t_0$, the exact transient is
\begin{align}
 u(t)
 &=
 u_{\rm ss}
 +
 e^{-\gamma\Delta t}
 \Big[
 (u_0-u_{\rm ss})\cos(2\omega\Delta t)
  \nonumber \\
& \qquad 
 +(v_0-v_{\rm ss})\sin(2\omega\Delta t)
 \Big],
 \label{eq:app-aniso-u-solution}
 \\
 v(t)
 &=
 v_{\rm ss}
 +
 e^{-\gamma\Delta t}
 \Big[
 -(u_0-u_{\rm ss})\sin(2\omega\Delta t)
\nonumber \\
& \qquad
 +(v_0-v_{\rm ss})\cos(2\omega\Delta t)
 \Big].
 \label{eq:app-aniso-v-solution}
\end{align}

Equation~\eqref{eq:aniso-residual} therefore contains, in general, both a
stationary contribution and a damped transient at the orientation frequency
$2\omega$.  The $2\omega$ component is a useful signature of anisotropic
diffusion during the transient, but it is not by itself universal: for example,
a state prepared directly in the anisotropic stationary covariance has only
the stationary contribution.

The distinction from the isotropic thermal generator is especially transparent
when the two dissipators have the same diffusion trace, 
$\bar d=\nu_{\rm th}.$
Nontrivial same-trace anisotropic choices satisfying this condition exist for
$\nu_{\rm th}>1/2$; at the zero-temperature limit
$\nu_{\rm th}=1/2$, complete positivity together with the same-trace
condition forces $d_1=d_2=1/2$.
Their trace equations are then identical to Eq.~\eqref{eq:s-eq}, so that the same initial value of $s$ produces the same energy trajectory.
The $q$ dynamics, however, satisfies
\begin{equation}
 \dot q+2\gamma q
 =
 \frac{2\gamma\nu_{\rm th}}{\omega}E(t)
 +
 \gamma(d_2-d_1)u(t).
 \label{eq:app-aniso-q-energy}
\end{equation}
Integrating gives
\begin{align}
 q(t)
 ={}&
 e^{-2\gamma(t-t_0)}q(t_0)
 +
 \frac{2\gamma\nu_{\rm th}}{\omega}
 \int_{t_0}^{t}
 e^{-2\gamma(t-\tau)}E(\tau)\,\dd\tau
 \nonumber\\
 &+
 \gamma(d_2-d_1)
 \int_{t_0}^{t}
 e^{-2\gamma(t-\tau)}u(\tau)\,\dd\tau .
 \label{eq:app-aniso-integral}
\end{align}
The last term is absent for the isotropic thermal generator.  Hence two
dissipators can produce exactly the same energy relaxation while predicting
different $q$ and entropy trajectories.

\section{Oscillator entropy rate and total entropy production}
\label{app:entropy-production}

The appearance of negative values of $\dot S$ in Fig.~\ref{fig:contours} is
consistent with the second law because $\dot S$ is the entropy rate of the
oscillator alone.  The total entropy production includes the entropy flow to the
thermal bath.

Let
\begin{equation}
 \rho_{\rm th}
 =
 \frac{e^{-H/T_{\rm bath}}}{\tr(e^{-H/T_{\rm bath}})}
 \label{eq:app-rhoth}
\end{equation}
be the stationary Gibbs state of the generator
\eqref{eq:master}. 
Define
$$
\mathcal L_{\rm th}\rho
=
\gamma(N+1)\mathcal D[a]\rho
+
\gamma N\mathcal D[a^\dagger]\rho.
$$
 For a thermal GKLS dissipator satisfying detailed balance,
Spohn's inequality \cite{Spohn1978} gives
\begin{equation}
 \dot\Sigma
=
-\tr\left[
(\mathcal L_{\rm th}\rho)
(\ln\rho-\ln\rho_{\rm th})
\right]\ge0
 \label{eq:app-spohn}
\end{equation}
The unitary part does not contribute to the von Neumann entropy rate, so
\begin{equation}
 \dot S=-\tr[(\mathcal L_{\rm th}\rho) \ln\rho].
\end{equation}
At fixed Hamiltonian the heat current from the bath is
\begin{equation}
 \dot Q
 =
 \tr[H \mathcal L_{\rm th}\rho]
 =
 \dot E
 =
 -\gamma(E-\omega\nu_{\rm th}).
 \label{eq:app-heat}
\end{equation}
Using
$\ln\rho_{\rm th}=-H/T_{\rm bath}-\ln Z$, Eq.~\eqref{eq:app-spohn} becomes
\begin{equation}
 \boxed{
 \dot\Sigma
 =
 \dot S-\frac{\dot Q}{T_{\rm bath}}
 \ge0.
 }
 \label{eq:app-entropy-balance}
\end{equation}

Thus $\dot S$ itself need not be nonnegative.  For example, an unsqueezed state
with $\nu>\nu_{\rm th}$ cools toward the bath and has
\begin{equation}
 \dot S
 =
 \gamma S'(\nu)(\nu_{\rm th}-\nu)<0,
\end{equation}
while Eq.~\eqref{eq:app-entropy-balance} still guarantees nonnegative total
entropy production.  The same distinction applies to squeezed states in the
negative-$\dot S$ portion of the physical region.

\section{Technical details of the driven invariant continuation}
\label{app:driven-details}

This Appendix collects the technical material behind the concise driven
extension of Sec.~\ref{sec:driven}.  None of the exact constant-frequency
results in Secs.~\ref{sec:moments} and \ref{sec:closure} depends on this
construction.

\subsection{Kernel normalization and Gaussian dictionary}

The Gaussian state is written as
\begin{equation}
 \rho=Z^{-1}\exp[-\mathcal I/T_0],
\end{equation}
where only the ratio $\mathcal I/T_0$ is physical.  Following the convention of
the main text, we fix the normalization by choosing $T_0$ as the temperature of
the thermal state of $H(0)$ with the same entropy as the initial Gaussian state:
\begin{equation}
 \nu(0)
 =
 \frac12\coth\frac{\omega(0)}{2T_0}.
 \label{eq:app-T0-definition}
\end{equation}
The driven kernel construction is stated for mixed initial Gaussian states with $T_0 > 0$; pure states may be approached as limiting cases.
For an initially unsqueezed state this is its ordinary initial temperature.
With this convention $\omega_{\mathcal I}$ has the same dimension as
$\omega$.

Williamson reduction of the quadratic kernel
\eqref{eq:quadratic-kernel} gives
\begin{align}
 \langle x^2\rangle
 &=
 \frac{\nu g_-}{\omega_{\mathcal I}},
 &
 \langle xp\rangle_{\rm s}
 &=
 -\frac{\nu g_0}{\omega_{\mathcal I}},
 \nonumber\\
 \langle p^2\rangle
 &=
 \frac{\nu g_+}{\omega_{\mathcal I}},
 &
 \nu
 &=
 \frac12\coth\frac{\omega_{\mathcal I}}{2T_0}.
 \label{eq:app-driven-covariance}
\end{align}
Differentiating the last relation,
\begin{equation}
 \dot\nu
 =
 -\frac{1}{T_0}
 \left(\nu^2-\frac14\right)
 \dot\omega_{\mathcal I}.
 \label{eq:app-chain-rule}
\end{equation}
Equivalently,
\begin{equation}
 \omega_{\mathcal I}
 =
 T_0S'(\nu).
 \label{eq:app-omegaI-entropy}
\end{equation}

Using the instantaneous dimensionless quadratures
$X=\sqrt{\omega}\,x$ and $P=p/\sqrt{\omega}$, one obtains
\begin{align}
 s&=\frac{2\nu\omega_{\rm eff}}{\omega},
 &
 E&=\nu\omega_{\rm eff},
 \label{eq:app-dictionary-energy}\\
 r^2
 &=
 \nu^2
 \left[
 \left(\frac{\omega_{\rm eff}}{\omega}\right)^2-1
 \right].
 \label{eq:app-dictionary-r}
\end{align}
Thus $\omega_{\rm eff}=E/\nu$ is the energetic scale associated with the
Gaussian deformation, and $\omega_{\rm eff}=\omega$ exactly for unsqueezed
states.

For fixed $\omega$, the covariance and kernel descriptions of the Gibbs
family are equivalent.  With $\rho$ in Eq.~\eqref{eq:rho-driven-main}, a Gibbs state~\eqref{eq:gibbs-manifold} of the oscillator at temperature $T$ may be written either as
$\sigma=\nu(T)\,\I$ or as $\mathcal I=\lambda H,$ with $
\lambda=\frac{T_0}{T}.$
Equivalently,
\begin{equation}
\boxed{
\begin{aligned}
\sigma=\nu\,\I
&\Longleftrightarrow
\mathcal I=\lambda H
\\
&\Longleftrightarrow
g_0=0,\qquad
g_-=\lambda,\qquad
g_+=\lambda\omega^2
\\
&\Longleftrightarrow
\omega_{\rm eff}=\omega
\Longleftrightarrow
r=0 ,
\end{aligned}}
\label{eq:app-kernel-gibbs-family}
\end{equation}
where
\begin{equation}
\omega_{\mathcal I}=\lambda\omega,
\qquad
\nu(T)
=
\frac12\coth\frac{\omega_{\mathcal I}}{2T_0}
=
\frac12\coth\frac{\omega}{2T}.
\label{eq:app-gibbs-nu}
\end{equation}

The bath selects one particular member of this family.  For a bath
temperature $T_{\rm bath}$,
\begin{equation}
\lambda=g:=\frac{T_0}{T_{\rm bath}},
\qquad
\mathcal I_{\rm target}=gH,
\label{eq:app-kernel-target}
\end{equation}
and therefore
\begin{equation}
\omega_{\mathcal I}^{\rm target}=g\omega,
\quad
\nu_{\rm th}
=
\frac12\coth\frac{g\omega}{2T_0}
=
\frac12\coth\frac{\omega}{2T_{\rm bath}} .
\label{eq:app-target-dictionary}
\end{equation}

\subsection{Component equations and invariant-scale relaxation}

Substituting Eq.~\eqref{eq:quadratic-kernel} into the local relaxation law
\eqref{eq:local-relaxation-law} gives
\begin{align}
 \dot g_-&=-2g_0-\alpha(g_--g),
 \label{eq:app-gminus}\\
 \dot g_0&=\omega^2g_--g_+-\alpha g_0,
 \label{eq:app-gzero}\\
 \dot g_+&=2\omega^2g_0-\alpha(g_+-g\omega^2).
 \label{eq:app-gplus}
\end{align}
Differentiating
$\omega_{\mathcal I}^2=g_-g_+-g_0^2$ and using these equations gives
\begin{equation}
 \dot\omega_{\mathcal I}
 =
 \alpha(g\omega_{\rm eff}-\omega_{\mathcal I}),
\end{equation}
which is Eq.~\eqref{eq:omegaI-relaxation}.  In particular,
$\alpha=0$ implies $\dot\omega_{\mathcal I}=0$ exactly, irrespective of the
Hamiltonian driving.

\subsection{Matching to the fixed-frequency GKLS benchmark}

The exact covariance dictionary does not imply that
Eq.~\eqref{eq:local-relaxation-law} is an exact rewriting of the benchmark
master equation.  At fixed frequency, Eq.~\eqref{eq:nu-eq} becomes
\begin{equation}
 \dot\nu
 =
 \gamma
 \left(
 \frac{\nu_{\rm th}\omega_{\rm eff}}{\omega}-\nu
 \right).
 \label{eq:app-GKLS-rate}
\end{equation}
On the other hand, combining Eqs.~\eqref{eq:app-chain-rule} and
\eqref{eq:omegaI-relaxation} gives
\begin{equation}
 \dot\nu
 =
 -\frac{\alpha}{T_0}
 \left(\nu^2-\frac14\right)
 (g\omega_{\rm eff}-\omega_{\mathcal I}).
\end{equation}
Equality of the entropy trajectories therefore requires
\begin{equation}
 \boxed{
 \alpha(g\omega_{\rm eff}-\omega_{\mathcal I})
 =
 \frac{T_0\gamma}{\nu^2-\tfrac14}
 \left(
 \nu-\frac{\nu_{\rm th}\omega_{\rm eff}}{\omega}
 \right).
 }
 \label{eq:app-alpha-gamma}
\end{equation}
Hence $\alpha$ should not in general be identified with the constant GKLS rate
$\gamma$.

On the Gibbs family, $\omega_{\rm eff}=\omega$, and
Eq.~\eqref{eq:app-alpha-gamma} reduces to
\begin{equation}
 \alpha
 =
 \frac{T_0\gamma(\nu-\nu_{\rm th})}
 {(\nu^2-\tfrac14)(g\omega-\omega_{\mathcal I})}.
 \label{eq:app-alpha-gibbs}
\end{equation}
Using
\begin{equation}
 \frac{\partial\omega_{\mathcal I}}{\partial\nu}
 =
 -\frac{T_0}{\nu^2-\tfrac14},
\end{equation}
one finds $\alpha\to\gamma$ near the bath target.  Matching the full covariance
trajectory away from that regime requires more than matching the entropy
equation alone.  This is why the driven law is described as a local continuation
rather than an exact change of variables.

\subsection{Moving Gibbs target and source of transverse excitation}

For fixed $\omega$, the kernel representation of the Gibbs family is
\begin{equation}
 \MG=
 \{\mathcal I=\lambda H\},
 \qquad
 g_0=0,
 \qquad
 g_+=\omega^2g_-.
 \label{eq:app-kernel-family}
\end{equation}
When $\omega(t)$ varies, the corresponding family moves with the Hamiltonian.
Defining
\begin{equation}
 \delta\mathcal I=\mathcal I-gH,
\end{equation}
Eq.~\eqref{eq:local-relaxation-law} gives
\begin{equation}
 \boxed{
 D_t\delta\mathcal I+\alpha\delta\mathcal I
 =
 -D_t(gH)
 =
 -\dot g\,H-g\dot H.
 }
 \label{eq:app-moving-target}
\end{equation}
The first source moves the target along the Gibbs direction.  The second
contains a transverse pair component.  With the instantaneous oscillator
operator $a_t$,
\begin{equation}
 \dot H
 =
 \frac{\dot\omega}{2}
 \left(
 a_t^2+a_t^{\dagger2}+2a_t^\dagger a_t+1
 \right).
 \label{eq:app-Hdot-pair}
\end{equation}
Thus varying $T_{\rm bath}$ at fixed $\omega$ does not by itself generate
squeezing, whereas varying $\omega$ generically does.  In this local
description the Hamiltonian drive generates the transverse deformation and the
bath relaxes it.

\subsection{Bath-correlation and secular conditions}

For a linear interaction proportional to $x$, the system operator in the
invariant basis has a matrix-element prefactor
\begin{equation}
 A(t)=\sqrt{\frac{g_-(t)}{2\omega_{\mathcal I}(t)}}.
 \label{eq:app-A}
\end{equation}
Define
\begin{equation}
 \Delta_{\mathcal I}
 =
 \frac{g\omega_{\rm eff}}{\omega_{\mathcal I}}-1.
\end{equation}
Then
\begin{equation}
 \frac{\dot\omega_{\mathcal I}}{\omega_{\mathcal I}}
 =
 \alpha\Delta_{\mathcal I},
 \qquad
 \tau_{\mathcal I}
 =
 \frac{1}{\alpha|\Delta_{\mathcal I}|}.
 \label{eq:app-tauI}
\end{equation}
A local Born--Markov construction requires the resonant data to vary
negligibly during the bath correlation time:
\begin{equation}
 \tau_B\ll
 \min(\alpha^{-1},\tau_{\mathcal I},\tau_A),
 \qquad
 \tau_A=\left|\frac{A}{\dot A}\right|.
 \label{eq:app-markov}
\end{equation}
The amplitude variation is
\begin{equation}
 \frac{\dot A}{A}
 =
 -\frac{g_0}{g_-}
 -\frac{\alpha}{2}
 \left[
 \frac{g_--g}{g_-}+\Delta_{\mathcal I}
 \right].
 \label{eq:app-Adot}
\end{equation}
Near the bath target in a weak-squeezing regime,
$|g_0/g_-|=O(\alpha)$, so $\tau_A=O(\alpha^{-1})$ and the condition reduces
parametrically to the ordinary weak-coupling Markov hierarchy.  Under strong
parametric driving, however, the $g_0/g_-$ term must be checked separately.

If a secular approximation is also imposed, the coarse-graining time must obey
\begin{equation}
 \max(\tau_B,\omega_{\mathcal I}^{-1})
 \ll\Delta t\ll
 \min(\alpha^{-1},\tau_{\mathcal I},\tau_A).
 \label{eq:app-secular}
\end{equation}
Outside this window a microscopic driven-bath treatment may require different
jump operators or a nonlocal memory kernel.


\end{document}